\documentclass[a4paper, final, 12pt]{article}
\usepackage{eurosym}
\usepackage{amsmath, amssymb, latexsym, amscd, amsthm,amsfonts,amstext}
\usepackage[mathscr]{eucal}
\usepackage{graphicx}
\usepackage{subfig}
\usepackage{float}
\usepackage{color}
\usepackage{hyperref}
\usepackage[utf8]{inputenc}
\usepackage[english]{babel}

\numberwithin{equation}{section}
\begin{document}

\title{Stationary multiple euclidon solutions to the vacuum Einstein equations}

\author{Aleksandr A. Shaideman$^{\ast }$ and
Kirill V. Golubnichiy$^{\circ }$ \\
 \and $^{\ast }$Department of Theoretical Physics
, \and  Peoples' Friendship University of Russia, \\ 
6 Miklukho-Maklaya St., 117198, Moscow,
 Russia \and $^{\circ }$Department of Mathematics \and Texas Tech University, Lubbock, TX 79409, USA  \and Emails: ashaideman@rambler.ru,
\and kgolubni@ttu.edu }
\date{}
\maketitle


\begin{abstract}
This paper is dedicated to Professor Ts.\ I.\ Gutsunaev, who  passed away.
In \cite{1}, we constructed the non-linear superposition of a stationary euclidon solution with an arbitrary axially symmetric stationary gravitational field using the method of variation of parameters. In the same work \cite{1}, the stationary soliton solution of the Einstein equations was generalized to the case of a stationary seed metric. The formulae obtained in \cite{1,2} have a simple, compact structure that enables an effective non-linear “addition” of solutions. In this framework, euclidon solutions serve as building blocks that generate almost all known vacuum, static, axially symmetric solutions of the Einstein equations, including the important Kerr–NUT family. Moreover, the stationary euclidon solution admits a clear physical interpretation as a relativistically accelerated, non-inertial reference frame, providing an alternative perspective on the physical meaning of celebrated solutions such as Kerr \cite{3}.

\end{abstract}


\vspace{0.5em} 





\section {Introduction}

     In \cite{1}, we constructed the non-linear superposition of a stationary euclidon solution with an arbitrary axially symmetric stationary gravitational field using the method of variation of parameters. In the same work \cite{1}, the stationary soliton solution of the Einstein equations was generalized to the case of a stationary seed metric. The formulae obtained in \cite{1,2} have a simple, compact structure that enables an effective non-linear “addition” of solutions. In this framework, euclidon solutions serve as building blocks that generate almost all known vacuum, static, axially symmetric solutions of the Einstein equations, including the important Kerr–NUT family. Moreover, the stationary euclidon solution admits a clear physical interpretation as a relativistically accelerated, non-inertial reference frame, providing an alternative perspective on the physical meaning of celebrated solutions such as Kerr \cite{3}.
     The first exact solutions of the axially symmetric stationary vacuum equations were presented by Lewis \cite{4} and Van Stockum \cite{5}. Later Papapetrou  \cite{6} obtained a canonical form of the line element, and Ernst \cite{7}  formulated the stationary vacuum problem in the form of a single equation for a complex function.                
     An outstanding result was obtained  in 1963 by Kerr \cite{8} , whose solution possibly describes the exterior gravitational field of a rotating source.
     A family of a asymptotically flat solutions of the stationary vacuum  equations was found by Tomimatsu and Sato \cite{9}, \cite{10}.
     Later, a survey of  known solutions was given by Kinnersley \cite{11} . Possible generalizations of the Tomimatsu-Sato metric were found by Yamazaki   \cite{12}, \cite{13} . Hori \cite{14} and Cosgrove \cite{15}, \cite{16}, and asymptotically not-flat solutions of the Ernst problem  were obtained in \cite{17}, \cite{19}, \cite{4}.
     The Ernst  equation \cite{7} makes it possible to extract  the solutions  by the method of separation of variables \cite{20}, \cite{21} . In a series of papers \cite{22}, \cite{23}, \cite{24}, a non-canonical form  of the stationary metrics  was used to obtain new solutions of the Einstein equations.
    An approximate method  for constructing solutions depending  on two harmonic functions was proposed  in \cite{25}, \cite{26}. Ernst \cite{27} dreamt of construction of such a kind of solutions, but exact ones. 
\section{Author Declarations}
    The authors have no conflicts to disclose.
\section{Basic equations}
As was shown  by Papapetrou \cite{6},  the line element, describing a stationary axially symmetric gravitational field without loss of generality can be present in the following canonical form:
\begin{equation}
ds^{2} =f^{-1}[e^{2\gamma}(d\rho^{2}+dz^{2})+\rho^{2}d\varphi^{2}]-f (dt-\omega d\varphi)^{2}\label{2.1}
\end{equation}
Here $\rho, \varphi, z$ and $t$ are canonical Weyl coordinates and time, respectively; $f(\rho, z)$,  $\gamma(\rho, z)$ and  $\omega(\rho, z)$ are  unknown functions to be determined from the  field equations. 

The  vacuum Einstein equations are given by

\begin{equation*}
R_{ik}=0,
\end{equation*}

where $R_{ik}$ is the  Ricci tensor .

Therefore we obtain all  the Einstein equations in the case  of a stationary axially symmetric gravitational field  outside  the sources:

\begin{equation}
f\Delta f =(\overrightarrow{\nabla}f)^{2}-f^{4}\rho^{-2}(\overrightarrow{\nabla}\omega)^{2}, \label{2.2}
\end{equation}

 \begin{equation}
\overrightarrow{\nabla} (f^{2}\rho^{-2}\overrightarrow{\nabla}\omega)=0, \label{2.3}
\end{equation}

  \begin{equation}
\frac{\partial \gamma}{\partial \rho}=\frac{1}{4}\rho f^{-2}[(\frac{\partial f}{\partial \rho})^{2}-(\frac{\partial f}{\partial z})^{2}-f^{4}\rho^{-2}((\frac{\partial \omega}{\partial \rho})^{2}-(\frac{\partial \omega}{\partial z})^{2})], \label{2.4}
\end{equation}

  \begin{equation}
 \frac{\partial \gamma}{\partial z}=\frac{1}{2} \rho f^{-2}(\frac{\partial f}{\partial \rho}\frac{\partial f}{\partial z}-f^{4}\rho^{-2}\frac{\partial \omega}{\partial \rho}\frac{\partial \omega}{\partial z}), \label{2.5}
\end{equation}
 
The operators  $\overrightarrow{\nabla}$  and $\Delta$  defined by the formulae

\begin{equation*}
\overrightarrow{\nabla}\equiv \overrightarrow{\rho}_{0}\frac{\partial}{\partial \rho}+\overrightarrow{z}_{0}\frac{\partial}{\partial z} \label{11}
\end{equation*}

\begin{equation*}
\Delta \equiv  \overrightarrow{\nabla}^{2} \equiv \frac{\partial^{2}}{\partial \rho ^{2}}+\frac{1}{\rho}\frac{\partial}{\partial \rho} +\frac{\partial^{2}}{\partial z^{2}}, \label{12}
\end{equation*}

( $\overrightarrow{\rho}_{0}$ and $\overrightarrow{z}_{0}$ being unit vectors), i.e., they are similar to the ordinary  Laplacian and gradient operators for flat space expressed in cylindrical coordinates provided that there is no angular coordinate dependence.

It should be noted that the integrability condition of Equations \ref{2.4}, \ref{2.5}  for determining the function  $\gamma(\rho, z)$  is the system \ref{2.2}, \ref{2.3} which does not contain    $\gamma(\rho, z)$ .  
Hence,  the problem of obtaining  exact axially symmetric solutions of  the Einstein Equations outside  the sources reduces to integration of the  differential Equations \ref{2.2}, \ref{2.3}

 The  prolate ellipsoidal coordinates $(x,y)$  are related to the Weyl canonical coordinates $(\rho,z)$  by the formulae 

\begin{equation*}
x=\frac{1}{2k_{0}}[\sqrt{\rho^{2}+(z+k_{0})^{2}}+\sqrt{\rho^{2}+(z-k_{0})^{2}}], 
\end{equation*}

 \begin{equation*}
y=\frac{1}{2k_{0}}[\sqrt{\rho^{2}+(z+k_{0})^{2}}-\sqrt{\rho^{2}+(z-k_{0})^{2}}], \label{13}
\end{equation*}

 where $k_{0}$  is a real constant. 
 
 The inverse transformation is

 \begin{equation*}
\rho=k_{0}\sqrt{(x^{2}-1)(1-y^{2})}, \quad z=k_{0}xy. \label{14}
\end{equation*}

In the prolate ellipsoidal coordinates $(x,y)$ the operators $\Delta$ and $\overrightarrow{\nabla}$  take the form 

\begin{equation*}
\Delta \equiv  \frac{1}{k^{2}_{0}(x^2-y^2)}(\frac{\partial}{\partial x}[(x^2-1)\frac{\partial}{\partial x}]+\frac{\partial}{\partial y}[(1-y^2)\frac{\partial}{\partial y}]),
\end{equation*}

\begin{equation*}
\overrightarrow{\nabla}\equiv  \frac{1}{k_{0}(x^2-y^2)^{1/2}}(\overrightarrow{x}_{0}\sqrt{x^2-1}\frac{\partial}{\partial x}+\overrightarrow{y}_{0}\sqrt{1-y^2}\frac{\partial}{\partial y}), 
\end{equation*}

$\overrightarrow{x}_{0}$ and $\overrightarrow{y}_{0}$  being unit vectors,

and Equations \ref{2.4}, \ref{2.5} take the form

\begin{equation*}
    \frac{\partial \gamma}{\partial x}=\frac{1-y^2}{4f^2(x^2-y^2)}\{x(1-y^2)[(\frac{\partial f}{\partial x})^2+(\frac{\partial \Phi}{\partial x})^2]-x(1-y^2)[(\frac{\partial f}{\partial y})^2+(\frac{\partial \Phi}{\partial y})^2]
\end{equation*}

\begin{equation}
    -2y(1-y^2)(\frac{\partial f}{\partial x}\frac{\partial f}{\partial y}+ \frac{\partial \Phi}{\partial x}\frac{\partial \Phi}{\partial y})\},\label{2.6}
\end{equation}

\begin{equation*}
    \frac{\partial \gamma}{\partial y}=\frac{x^2-1}{4f^2(x^2-y^2)}\{y(1-y^2)[(\frac{\partial f}{\partial x})^2+(\frac{\partial \Phi}{\partial x})^2]-y(1-y^2)[(\frac{\partial f}{\partial y})^2+(\frac{\partial \Phi}{\partial y})^2], 
\end{equation*}

\begin{equation}
    -2x(1-y^2)(\frac{\partial f}{\partial x}\frac{\partial f}{\partial y}+ \frac{\partial \Phi}{\partial x}\frac{\partial \Phi}{\partial y})\}. \label{2.7}
\end{equation}

The line element \ref{2.1} in this case can be rewritten as,
 
\begin{equation*}
ds^{2} =k^{2}_{0}f^{-1}[e^{2\gamma}(x^2-y^2)(\frac{ dx^2}{x^2-1}+\frac{dy^2}{1-y^2})+
\end{equation*}

\begin{equation}
+(x^2-1)(1-y^2)d\varphi^{2}]-f(dt-\omega d\varphi)^{2}, \label{2.8}
\end{equation}
where the metric coefficient $f$ , $\gamma$  and $\omega$ are functions of $x$ and $y$ only.

 Accordingly in this case  the Equation

\begin{equation}
\Delta \psi\equiv \frac{\partial^{2} \psi}{\partial \rho ^{2}}+\frac{1}{\rho}\frac{\partial \psi}{\partial \rho} +\frac{\partial^{2} \psi}{\partial z^{2}}=0 \label{2.9}
\end{equation}
becomes 

\begin{equation}
\frac{\partial}{\partial x}[(x^2-1)\frac{\partial \psi}{\partial x}]+\frac{\partial}{\partial y}[(1-y^2)\frac{\partial \psi}{\partial y}]=0, \label{2.10}
\end{equation}
and  the Equation

\begin{equation}
    \Delta' \chi \equiv \frac{\partial^2\chi}{\partial \rho^2}-\frac{1}{\rho}\frac{\partial\chi}{\partial \rho}+\frac{\partial^2\chi}{\partial z^2}=0 \label{2.11}
\end{equation}

can be rewritten as 

\begin{equation}
    (x^2-1)\frac{\partial^2\chi}{\partial x^2}+(1-y^2)\frac{\partial ^2 \chi}{\partial y^2}=0. \label{2.12}
\end{equation}

There exist two possibilities of writing down Equations \ref{2.2}, \ref{2.3}  and  in a coordinate-free form.

\textbf{1.} If we switch from the rotation potential $\omega$ to the new potential  $\Phi$  by the formulae 

\begin{equation}
    \frac{\partial \omega}{\partial \rho}=\frac{\rho}{f^2}\frac{\partial \Phi}{\partial z}, \quad \frac{\partial \omega}{\partial z}=-\frac{\rho}{f^2}\frac{\partial \Phi}{\partial \rho}, \label{2.13}
\end{equation}

or, in the prolate ellipsoidal coordinates, 
 
 \begin{equation}
    \frac{\partial \omega}{\partial x}=\frac{k_{0}(1-y^2)}{f^2}\frac{\partial \Phi}{\partial y}, \quad \frac{\partial \omega}{\partial y}=-\frac{k_{0}(x^2-1)}{f^2}\frac{\partial \Phi}{\partial x}, \label{2.14}
\end{equation}

we obtain the following field equations:

\begin{equation}
    f\Delta f=(\overrightarrow{\nabla}f)^2-(\overrightarrow{\nabla}\Phi)^2, \quad \overrightarrow{\nabla}(f^{-2}\overrightarrow{\nabla}\Phi)=0. \label{2.15}
\end{equation}

The set of  two second-order differential equations can be reduced to four first-order differential equations. 

Let us introduce  the notations

\begin{equation}
    f^{-1}\overrightarrow{\nabla}f=\overrightarrow{A}, \quad f^{-1}\overrightarrow{\nabla}\Phi=\overrightarrow{B}. \label{2.16}
\end{equation}

In this case,  from we obtain the following set of four first-order differential equations for $\overrightarrow{A}$ and $\overrightarrow{B}$:

\begin{equation*}
    div \overrightarrow{A}=-\overrightarrow{B}^2, \quad div\overrightarrow{B}=(\overrightarrow{A}\cdot \overrightarrow{B}),
\end{equation*}

\begin{equation}
    rot \overrightarrow{A}=0, \quad rot\overrightarrow{B}=-[\overrightarrow{A}\times \overrightarrow{B}], \label{2.17}
\end{equation}

This form of the field equations was developed in \cite{28}, \cite{29}.
In the Weyl canonical coordinates we can rewrite \ref{2.16} as

\begin{equation*}
    A_{1}=\frac{1}{f}\frac{\partial f}{\partial \rho}, \quad A_{2}=\frac{1}{f}\frac{\partial f}{\partial z},
\end{equation*}

\begin{equation}
    B_{1}=\frac{1}{f}\frac{\partial \Phi}{\partial \rho}, \quad B_{2}=\frac{1}{f}\frac{\partial \Phi}{\partial z}, \label{2.18}
\end{equation}

In this case, using \ref{2.17} and \ref{2.18}, we obtain

\begin{equation*}
    \frac{1}{\rho}\frac{\partial}{\partial \rho}(\rho A_{1})+\frac{\partial A_{2}}{\partial z}=-(B^2_{1}+B^2_{2}),
\end{equation*}

\begin{equation}
    \frac{1}{\rho}\frac{\partial}{\partial \rho}(\rho B_{1})+\frac{\partial B_{2}}{\partial z}=A_{1}B_{1}+A_{2}B_{2}, \label{2.19}
\end{equation}

\begin{equation*}
    \frac{\partial A_{1}}{\partial z}-\frac{\partial A_{2}}{\partial \rho}=0, \quad  \frac{\partial B_{1}}{\partial z}-\frac{\partial B_{2}}{\partial \rho}=A_{2}B_{1}-A_{1}B_{2}.
\end{equation*}

Let us  consider an analogue of the system \ref{2.19} in the ellipsoidal coordinates $(x,y)$.  

Introducing the functions 

\begin{equation*}
    \alpha_{1}=\frac{1}{f}\frac{\partial f}{\partial x}, \quad \alpha_{2}=\frac{1}{f}\frac{\partial f}{\partial y},
\end{equation*}

\begin{equation}
    \beta_{1}=\frac{1}{f}\frac{\partial \Phi}{\partial x}, \quad \beta_{2}=\frac{1}{f}\frac{\partial \Phi}{\partial y}, \label{2.20}
\end{equation}

we  have from Equations \ref{2.19}  the following equivalent set of four first-order differential equations:

\begin{equation*}
    \frac{\partial}{\partial x}[(x^2-1)\alpha_{1}]+\frac{\partial}{\partial y}[(1-y^2)\alpha_{2}]=-(x^2-1)\beta^2_{1}-(1-y^2)\beta^2_{2},
\end{equation*}

\begin{equation}
    \frac{\partial}{\partial x}[(1-y^2)\beta_{1}]+\frac{\partial}{\partial y}[(1-y^2)\beta_{2}]=-(1-y^2)\alpha_{1}\beta_{1}+(1-y^2)\alpha_{2}\beta_{2}, \label{2.21}
\end{equation}

\begin{equation*}
    \frac{\partial \alpha_{2}}{\partial x}-\frac{\partial \alpha_{1}}{\partial y}=0, \quad \frac{\partial \beta_{2}}{\partial x}-\frac{\partial \beta_{1}}{\partial y}=\alpha_{2}\beta_{1}-\alpha_{1}\beta_{2}.
\end{equation*}

Introducing the Ernst complex potential 

\begin{equation}
    \varepsilon=f+i\Phi \label{2.22}
\end{equation}

gives us, from \ref{2.15},  the equation 

\begin{equation}
    (\varepsilon+\varepsilon^{*})\Delta \varepsilon=2(\overrightarrow{\nabla}\varepsilon)^2,\label{2.23}
\end{equation}

where $\varepsilon^{*}$  is the complex conjugate of  $\varepsilon$.

A transformation of \ref{2.23}

\begin{equation}
    \varepsilon=(\xi-1)(\xi+1)^{-1},\label{2.24}
\end{equation}

leads  to one more form of  the  equations for a stationary axially symmetric gravitational  field:

\begin{equation}
    (\xi\xi^{*}-1)\Delta \xi=2\xi^{*}(\overrightarrow{\nabla}\xi)^2.\label{2.25}
\end{equation}

For  Equation \ref{2.23}  we can prove the following theorem :

\textbf{Theorem 1.} \emph{If \ \  $\Tilde{\varepsilon}$ is a solution of equation, then the functions} 
\begin{equation}
\varepsilon=\frac{iA_{0}+B_{0}\Tilde{\varepsilon}}{C_{0}+iD_{0}\Tilde{\varepsilon}}, \quad \varepsilon^{*}=\frac{-iA_{0}+B_{0}\Tilde{\varepsilon^{*}}}{C_{0}-iD_{0}\Tilde{\varepsilon^{*}}},\label{2.26}
\end{equation}

\emph{where $A_{0}, B_{0}, C_{0}$ and $D_{0}$ are arbitrary real constants subject to the constraint  $A_{0}D_{0}+B_{0}C_{0}\neq 0,$ also satisfy Equation \ref{2.23}.}

Its proof  can be obtained immediately by substitution of  into the  corresponding Equation in \ref{2.23}.

From Equations \ref{2.26} ,  with the help of \ref{2.22}  , one can obtain:

\begin{equation*}
    f=\frac{(A_{0}D_{0}+B_{0}C_{0})\Tilde{f}}{(C_{0}-D_{0}\Tilde{\Phi})^{2}+D^2_{0}\Tilde{f}^2},
\end{equation*}

\begin{equation}
    \Phi=\frac{A_{0}C_{0}+(B_{0}C_{0}-A_{0}D_{0})\Tilde{\Phi}-B_{0}D_{0}(\Tilde{f}^2+\Tilde{\Phi}^2)}{(C_{0}-D_{0}\Tilde{\Phi})^2+D^2_{0}\Tilde{f}^2}+E_{0},\label{2.27}
\end{equation}

where  $\Tilde{f} $   and   $\Tilde{\Phi} $  satisfy Equations \ref{2.15}. It should be noted that if we let in  $A_{0}=E_{0}=0,\ \ B_{0}=C_{0}=1$, we obtain the Ehlers transformation ( \cite{37}, see also \cite{11} )

\begin{equation}
    f=\frac{\Tilde{f}}{(1-D_{0}\Tilde{\Phi})^2+D^2_{0}\Tilde{f}^2}, \quad \Phi=\frac{\Tilde{\Phi}-D_{0}(\Tilde{f}^2+\Tilde{\Phi}^2)}{(1-D_{0}\Tilde{\Phi})^2+D^2_{0}\Tilde{f}^2}. \label{2.28}
\end{equation}

The inverse transformation 

\begin{equation}
    \varepsilon=\frac{1}{\Tilde{\varepsilon}}, \quad \varepsilon^{*}=\frac{1}{\Tilde{\varepsilon^{*}}}. \label{2.29}
\end{equation}

is recognized to be a special case of  \ref{2.26} corresponding to the choice  $ B_{0}=C_{0}=E_{0}=0,\ \ A_{0}=D_{0}$.
So \ref{2.27} take the form 

\begin{equation}
    f=\frac{\Tilde{f}}{\Tilde{f}^2+\Tilde{\Phi}^2}, \quad \Phi=\frac{\Tilde{\Phi}}{\Tilde{f}^2+\Tilde{\Phi}^2}.\label{2.30}
\end{equation}

It can be pointed out that the inverse transformation leaves the metric function  $\gamma$ unchanged, i.e.,  $\gamma = \Tilde{\gamma}$.

As an another particular case of \ref{2.26}, the  constant phase transformation 

\begin{equation}
    \xi=e^{i\alpha_{0}}\Tilde{\xi},\label{2.31}
\end{equation}

where $\alpha_{0}$ is real constant and the function $\Tilde{\xi}$ is introduced according \ref{2.24}, gives us a possibility of obtaining new solutions of the  Ernst equation \ref{2.25} from old ones.

The transformation 

\begin{equation}
    \varphi \rightarrow C_{1}\varphi+C_{2}t, \quad t\rightarrow C_{3}\varphi+C_{4}t,\label{2.32}
\end{equation}

where $C_{1},\ \ C_{2},\ \ C_{3}$ and $C_{4}$ are real constants, preserves the characteristic form of the Papapetrou line element \ref{2.1}. In this connection we can proof the following theorem.

\textbf{2.} The substitution $f=\rho / F$  in  equations gives us the following  equations:

\begin{equation}
    F\Delta F=(\overrightarrow{\nabla}F)^2+(\overrightarrow{\nabla}\omega)^2, \quad\overrightarrow{\nabla}(F^{-2}\overrightarrow{\nabla}\omega)=0.\label{2.33}
\end{equation}

The set of two second-order differential equations can also be reduced to a set of four first-order differential equations.

If we introduce the notations 

\begin{equation}
    F^{-1}\overrightarrow{\nabla}F=A, \quad F^{-1}\overrightarrow{\nabla}\omega=B,\label{2.34}
\end{equation}

we obtain (\ref{2.17}).

Introducing the functions 

\begin{equation}
    \varepsilon_{1}=F+\omega \quad \varepsilon_{2}=F-\omega,\label{2.35}
\end{equation}

we rewrite Equations (\ref{2.33}) in the symmetric form

\begin{equation*}
    (\varepsilon_{1}+\varepsilon_{2})\Delta \varepsilon_{1}=2(\overrightarrow{\nabla}\varepsilon_{1})^2,
\end{equation*}
\begin{equation}
    (\varepsilon_{1}+\varepsilon_{2})\Delta \varepsilon_{2}=2(\overrightarrow{\nabla}\varepsilon_{2})^2.\label{2.36}
\end{equation}
For the field equations and one can prove theorems, similar to and. So, for equations we have the following theorem.

\textbf{Theorem 2.} \emph{If  $\Tilde{\varepsilon_{1}}$ and $\Tilde{\varepsilon_{2}}$      are solution of  equations , then the functions} 

\begin{equation}
    \varepsilon=\frac{A_{0}+B_{0}\Tilde{\varepsilon}_{1}}{C_{0}+D_{0}\Tilde{\varepsilon}_{1}}, \quad \varepsilon_{2}=\frac{-A_{0}+B_{0}\Tilde{\varepsilon}_{2}}{C_{0}-D_{0}\Tilde{\varepsilon}_{2}}, \label{2.37}
\end{equation}

\emph{where $A_{0}, B_{0}, C_{0}$ and $D_{0}$ are arbitrary real constants subject to the constraint  $B_{0}C_{0} -A_{0}D_{0}\neq 0,$ also satisfy equations.} 

From \ref{2.37} with the help of \ref{2.35} we obtain

\begin{equation*}
    F=\frac{(B_{0}C_{0}-A_{0}D_{0})\Tilde{F}}{(C_{0}+D_{0}\Tilde{\omega})^{2}-D^2_{0}\Tilde{F}^2},
\end{equation*}

\begin{equation}
    \omega=\frac{A_{0}C_{0}+(A_{0}D_{0}+B_{0}C_{0})\Tilde{\omega}-B_{0}D_{0}(\Tilde{F}^2-\Tilde{\omega}^2)}{(C_{0}+D_{0}\Tilde{\omega})^2-D^2_{0}\Tilde{F}^2}, \label{2.38}
\end{equation}

where $\Tilde{F}$ and $\Tilde{\omega}$  satisfy Equation \ref{2.23}.

The papers \cite{30}-\cite{33} were devoted to the symmetric of vacuum stationary axially symmetric Einstein equations.

One more possible formulation of the axially symmetric equations was proposed in 1978 by Cosgrove \cite{38}, \cite{39}. Rather than examine the system of two  second-order partial differential equations for the function $f$ and $\Phi$, Cosgrove prefered to analyze one  fourth-order partial differential equation for the metric function $\gamma$.

Finally, it should be noted that the procedure for calculating the relativistic multipole moments was given by Geroch \cite{40}, \cite{41} and Hansen \cite{42}.

A simplified procedure for calculating the Geroch and Hansen multipole moments, was proposed by Hoenselaers \cite{43}, who found a relation between the Ernst potential of an arbitrary axially symmetric stationary vacuum metric and the corresponding Geroch-Hansen  multipole moments.

The relativistic and coordinate-invariant definitions of multipole moments were also proposed in \cite{44}, \cite{45}. Although one is led to these definitions by different mathematical approaches, it can be shown that all of them are physically equivalent \cite{46}, \cite{47}.

\section{The Lewis's class of solutions. One-stationary euclidon solution}

In 1932 Lewis \cite{4} obtained the following class of exact solutions of Equations (ref{2.13}, \ref{2.15})  and:

\begin{equation*}
    f_{L}=\frac{\rho}{C_{1}}\frac{\sinh(\psi+U_{0})}{\cosh U_{0}},
\end{equation*}

\begin{equation}
    \Phi_{L}=\frac{1}{C_{1}\cosh U_{0}}\frac{\partial\chi}{\partial z}+C_{2}, \label{2.39}
\end{equation}

\begin{equation*}
    \omega_{L}=C_{1}\frac{\cosh \psi}{\sinh(\psi+U_{0})}+C_{3}.
\end{equation*}

Here $C_{1}$, $C_{2}$, $C_{3}$ and $U_{0}$ are arbitrary constants, $\psi(\rho,z)=(1/ \rho)(\partial\chi/\partial\rho)$ and $\chi(\rho,z)$ are functions satisfying the linear Equations   \ref{2.9},  \ref{2.11}.

Note that Weyl's static solution $f_{\mathcal{W}}=(\rho/C_{1})e^{-\psi}, \Phi_{\mathcal{W}}=C_{2}$ and $\omega_{\mathcal{W}}=C_{3}$ (where $C_{2}$ and $C_{3}$ are usually set equal to zero) can be obtained from  \ref{2.39} in the limit $U_{0}\rightarrow \infty.$

To obtain the basic classes of stationary vacuum solutions in a more or less unified way, we shall use the representation of Equations \ref{2.15}  in the form of four first-order differential equations  \ref{2.19}  for the functions  \ref{2.18} , which we shall further subject to certain algebraic dependences.

The class of solutions  \ref{2.39} can be obtained by making the functions  \ref{2.18}satisfy the condition

\begin{equation}
    A_{1}B_{1}+A_{2}B_{2}=\frac{B_{1}}{\rho}. \label{2.40}
\end{equation}

Then one can introduce the new function $M(\rho, z)$ such that

\begin{equation}
    B_{1}=MA_{2}, \quad B_{2}=-M(A_{1}-\frac{1}{\rho}).\label{47}
\end{equation}

Then from \ref{2.19} we have the equation 

\begin{equation}
    M(M^2-1)\Delta M=(2M^2-1)(\overrightarrow{\nabla}M)^2.\label{2.42}
\end{equation}

If the function $M(\rho, z)$ is known, one can easily find the functions $F$ and $\Phi$ from Equations \ref{2.18} with help of the relations

\begin{equation*}
    A_{1}=\frac{1}{M(M^2-1)}\frac{\partial M}{\partial \rho}+\frac{1}{\rho}, \quad A_{2}=\frac{1}{M(M^2-1)}\frac{\partial M}{\partial z},
\end{equation*}

\begin{equation}
    B_{1}=\frac{1}{M^2-1}\frac{\partial M}{\partial z}, \quad B_{2}=\frac{1}{1-M^2}\frac{\partial M}{\partial \rho}. \label{2.43}
\end{equation}

\textbf{1.} Let us consider the most important cases of the class of solutions under study.

\textbf{1-I.} $M^2>1$. In this case Equation \ref{2.42} , by the substitution 

\begin{equation}
    M=\frac{1}{\cos(\psi+U_{0})}, \label{2.44}
\end{equation}

reduces to the linear Equation \ref{2.9} . In this case we obtain the solutions 

\begin{equation*}
    f=\frac{\rho}{\cos U_{0}}\sin(\psi+U_{0}), \quad \Phi=\frac{1}{\cos U_{0}}\frac{\partial \chi}{\partial z},
\end{equation*}

\begin{equation}
    \omega=\frac{\cos \psi}{\sin(\psi+U_{0})}, \label{2.45}
\end{equation}

\begin{equation*}
    \gamma=\frac{1}{2}\ln(\sin\psi+\frac{\cos\psi}{\cos U_{0}})+\frac{1}{4}\ln\rho +\gamma',
\end{equation*}

where the functions $\psi$ and $\chi$ satisfy Equations \ref{2.9} and \ref{2.11}, respectively, and are related to each other by $\psi=(1/\rho)(\partial \chi/\partial \rho).$ The function $\gamma'$ can be obtained from Equations.

The function $\gamma'$ is defined by 

\begin{equation}
    \frac{\partial \gamma'}{\partial \rho}=\frac{\rho}{4}[(\frac{\partial \psi}{\partial \rho})^2-(\frac{\partial \psi}{\partial z})^2], \quad \frac{\partial \gamma'}{\partial z}=\frac{\rho}{2}\frac{\partial \psi}{\partial \rho}\cdot \frac{\partial \psi}{\partial z}. \label{2.46}
\end{equation}

It should be noted that the solution \ref{2.45} can be transformed to the static vacuum Weyl class by a complex linear transformation of the coordinates $\varphi$ and $t$ (\cite{55}).

\textbf{1-II.} $M^2=1$.

The solution in this case has the form 

\begin{equation*}
    f=\frac{\partial \chi}{\partial \rho}, \quad \Phi=\frac{\partial \chi}{\partial z},
\end{equation*}

\begin{equation}
    \omega=-\rho (\frac{\partial\chi}{\partial \rho})^{-1}, \quad \gamma=\frac{1}{2}\ln\frac{\partial \chi}{\partial \rho}-\frac{1}{4}\ln\rho, \label{2.47}
\end{equation}

\begin{equation*}
    \Delta'\chi=0,
\end{equation*}

and it determines the Van Stockum metric \cite{5}.

\textbf{1-III.} $M^2<1$. In this case

\begin{equation}
    M=\cosh^{-1}(\psi+U_{0}) \quad (U_{0}=const), \label{2.48}
\end{equation}

and we obtain \ref{2.39}.

The metric function  $\gamma$ is in this case 

\begin{equation}
    \gamma= \gamma_{L}=\frac{1}{2}\ln(\sinh \psi+\frac{\cosh \psi}{\cosh U_{0}})+\frac{1}{4}\ln\rho+\gamma', \label{2.49}
\end{equation}

The function $\gamma'$ can be obtained from \ref{2.46}.

\ref{2.39} determine the static vacuum Weyl class because the function $\omega$ can be eliminated by a linear transformation  

\begin{equation*}
    \varphi\rightarrow C^{0}_{1}\varphi+C^{0}_{2}t, \quad t\rightarrow C^{0}_{3}\varphi+C^{0}_{4}t,
\end{equation*}

where $C^{0}_{1}, C^{0}_{2}, C^{0}_{3}$ and $C^{0}_{4}$ are real constants. 

\textbf{(a)} If one takes the expression

\begin{equation*}
    \chi=\chi_{1}=\int\sqrt{\rho^2+(z-z_{1]})^2}dz, 
\end{equation*}
 then 

\begin{equation*}
    \psi=\psi_{1}=\frac{1}{\rho}\frac{\partial \chi_{1}}{\partial \rho}=-\ln\rho+\ln[(z-z_{1}+\sqrt{\rho^2+(z-z_{1})^2})],
\end{equation*}

where $z_{1}$ is a displacement constant.

The solution in this particular case has the form

\begin{equation*}
    f=f_{1}=(z-z_{1})+\sqrt{\rho^2+(z-z_{1})^2}\cdot \tanh U_{0},
\end{equation*}
\begin{equation}
    \Phi=\Phi_{1}=\frac{\sqrt{\rho^2+(z-z_{1})^2}}{\cosh U_{0}}, \quad \omega=\omega_{1}=\frac{\sqrt{\rho^2+(z-z_{1})^2}}{\cosh U_{0}}\cdot\frac{1}{f_{1}},\label{2.50}
\end{equation}

\begin{equation*}
   \gamma=\gamma_{1}=\frac{1}{2}\ln f_{1}-\frac{1}{2}\ln\sqrt{\rho^2+(z-z_{1})^2}. 
\end{equation*}

The line element then takes the form

\begin{equation}
    ds^2=\frac{d\rho^2+dz^2}{\sqrt{\rho^2+(z-z_{1})^2}}+\frac{\rho^2}{f_{1}}d\varphi^2-f_{1}[dt-\frac{\sqrt{\rho^2+(z-z_{1})^2}}{\cosh U_{0}}\cdot \frac{1}{f_{1}}d\varphi]^2. \label{2.51}
\end{equation}

It can be checked that a direct substitution of the solution \ref{2.50} nullifies all components of the Riemann tensor, which corresponds to flat space. 

So we call the solution a stationary euclidon.

\textbf{(b)} If we choose the function $\chi$ in the form 

\begin{equation*}
    \chi=\chi_{2}=\frac{z^2}{2}-\frac{\rho^2}{4}(2\ln \rho-1),
\end{equation*}

then  we obtain the solutions

\begin{equation*}
    f=f_{2}=W(\rho), \quad \Phi=\Phi_{2}=\frac{z}{\cosh U_{0}},
\end{equation*}

\begin{equation}
    \omega=\omega_{2}=\frac{1+\rho^2}{\rho\cosh U_{0}}\frac{1}{W(\rho)}, \label{2.52}
\end{equation}

\begin{equation*}
    \gamma=\gamma_{2}=\frac{\ln W(\rho)}{2}, \quad W(\rho)=\frac{1+\tanh U_{0}}{2}-\frac{1-\tanh U_{0}}{2}\cdot\rho^2,
\end{equation*}

which is a special case of the general stationary cylindrically symmetric solution \cite{48}, i.e., the Kasner metric ( \cite{49}, see also \cite{50}-\cite{54} ). The line element in this case takes the form

\begin{equation}
    ds^2=d\rho^2+dz^2+\frac{\rho^2}{W(\rho)}d\varphi^2-W(\rho)[dt-\frac{1+\rho^2}{\rho\cosh U_{0}}\frac{1}{W(\rho)}d\varphi]^2,\label{2.53}
\end{equation}

which can be obtained from the Minkowski metric by transformation 

\begin{equation}
    \varphi'=\frac{1+\tanh U_{0}}{2}\varphi+\frac{1-\tanh U_{0}}{2}t, \quad t'=\frac{1+\tanh U_{0}}{2}t. \label{2.54}
\end{equation}

Therefore, we call the solution a stationary euclidon of second kind.

\textbf{2.} Let us finally consider simple methods of solving Equations

\textbf{2-I.} We rewrite Equations \ref{2.33} in the form

\begin{equation}
  \overrightarrow{\nabla}[\frac{\overrightarrow{\nabla}(F^2-\omega^2)}{F^2}]=0, \quad \overrightarrow{\nabla}(\frac{\overrightarrow{\nabla}\omega}{F^2})=0.\label{2.55}  
\end{equation}

and choose the solution of Equations in the form $F=F(\psi)$ and $\omega=\omega (\psi)$ where $\psi$ is an arbitrary solution of the equation $\Delta \psi =0$.

In this case, integrating the set of ordinary differential equations

\begin{equation}
    \frac{d}{d \psi}[\frac{1}{F^2}\frac{d}{d \psi}(F^2-\omega^2)]=0, \quad \frac{d}{d \psi}(\frac{1}{F^2}\frac{d \omega}{d\psi})=0, \label{2.56}
\end{equation}

we find 

\begin{equation}
    \frac{d}{d\psi}(F^2-\omega^2)=a_{0}F^2, \quad \frac{d\omega}{d\psi}=b_{0}F^2, \label{2.57}
\end{equation}

\begin{equation*}
    a_{0}=const., \quad b_{0}=const. 
\end{equation*}
The integration \ref{2.57}, as well as the following choice and renaming of the constants, 

\begin{equation*}
    b_{0}=-\frac{1}{C_{1}\cosh U_{0}}, \quad a_{0}=\frac{2C_{3}}{C_{1}\cosh U_{0}}-2\tanh U_{0}, 
\end{equation*}

brings the solution to its final form 

\begin{equation*}
    f=\rho\frac{\sinh(\psi+U_{0})}{C_{1}\cosh U_{0}}, \quad \omega=\frac{C_{1}\cosh\psi}{\sinh(\psi+U_{0})}+C_{3},
\end{equation*}

i.e., to the Lewis class of solutions.

\textbf{2-II.} The Lewis class of solutions can be easily found with the aid of the symmetry transformation.

In a special case, setting in \ref{2.38}

\begin{equation*}
    \Tilde{F}=e^{-\psi} \ \ (\Delta\psi=0, U_{0}=const.), 
\end{equation*}

\begin{equation*}
    \Tilde{\omega}=0, \ \ A_{0}=-D_{0}=e^{-U_{0}}, B_{0}=C_{0}=1,
\end{equation*}

we obtain 

\begin{equation*}
    F=\frac{\cosh U_{0}}{\sinh(\psi+U_{0})}, \quad \omega=\frac{\cosh \psi}{\sinh(\psi+U_{0})}.
\end{equation*}

Recalling that $F=\rho / f$, we arrive at the Lewis class of solutions.

\section{The euclidon method}

A study of methods of generating exact solutions, based on internal symmetries of the Einstein equations, has been attracting the attention of gravitation physicists. 

The publications by Ehlers \cite{56} , Ozvath \cite{57} and Harrison \cite{58} pioneer these investigations. Subsequently, one can distinguish three approaches to this problem. 

The group-theoretic technique, which is very useful in construction of the new metrics containing an arbitrary large number of parameters, was introduced by Geroch \cite{59}, Kinnersley \cite{60} and Maison \cite{61}, and used in the papers \cite{62}-\cite{75}. The main achievements of this approach are related with the group of continuous transformations of the Ernst equation, with the Hoenselaers-Kinnersley-Xanthopoulos (HKX) transformations \cite{68},\cite{69}, with the aid of which one can obtain a set of asymptotically flat stationary vacuum metrics \cite{76}-\cite{98}. 

The second approach, introduced originally by Belinsky and Zakharov \cite{99}, 
 \cite{100}, is based on the application of the inverse scattering method to the Einstein's equations. At this point one can obtain new results in general relativity \cite{101}-\cite{120} . 

Finally, the last approach to generate new solutions from old was founded on the applications of the Backlund transformations. For the first time in general relativity the Backlund transformations were introduced by Harrison \cite{121} and Neugebauer \cite{122}, and their possible usage one can found from \cite{123}-\cite{131}.

Although these approaches are independent, they all are mathematically equivalent, that was shown by Cosgrove \cite{132}-\cite{134}. 

The major result of the applications of techniques discussed above is the construction of the non-linear superposition of N Kerr solutions aligned along their common rotational axis. Therewith the Tomimatsu-Sato metric has been recognized to be a special case of this solution \cite{135}-\cite{155}. 

Furthermore, the stationary vacuum problem can be reduced to the one singular equation \cite{156}. The application of the technique of the generalized Riemannian problem to general relativity yields a large class of new solutions \cite{157}.

In the paper \cite{158} the method of the variation of constants was proposed. With
the help of this technique one can brought to the nonlinear superposition of the Kerr space-time, with an arbitrary vacuum field \cite{159}, \cite{160}. 

In \cite{1} the nonlinear superposition of the stationary euclidon solution with an arbitrary axially symmetric stationary gravitational field on the basis of the method of variation of parameters was constructed. In the paper 
  \cite{2} stationary soliton solution of the Einstein equations was generalized to the case of a stationary seed metric. The formulae obtained in \cite{1}, \cite{2}  have a simple and compact form, permitting an effective nonlinear "addition" of the solutions. The euclidon method propounded in \cite{1}  was used in the article \cite{161}. 

In this paper the both methods \cite{1}, \cite{2}  are considered again and some new applications are presented. 

Let $f^0(\rho, z)$,  $\Phi^0(\rho,z)$, $\omega^0(\rho,z)$ be some arbitrary solution of the equations 

\begin{equation*}
    f\Delta f= (\overrightarrow{\nabla}f)^2 -(\overrightarrow{\nabla}\Phi)^2, \quad f\Delta \Phi=2\overrightarrow{\nabla}\Phi \overrightarrow{\nabla}f,
\end{equation*}

\begin{equation}
    \frac{\partial \omega}{\partial \rho}=\frac{\rho}{f^2}\frac{\partial \Phi}{\partial z}, \quad \frac{\partial \omega}{\partial z}=-\frac{\rho}{f^2}\frac{\partial \Phi}{\partial \rho}. \label{2.58}
\end{equation}

 ($f^0,\ \ \Phi^0,\ \ \omega^0$) - we call the seed solution.

 One of the simplest solutions of Equation  (\ref{2.58}) is the one-stationary euclidon solution

\begin{equation*}
    f_E = \frac{(z - z_1) + \sqrt{\rho^2 + (z - z_1)^2} \tanh U_0}{C_{1}},
\end{equation*}

\begin{equation}
    \Phi_E = \frac{\sqrt{\rho^2 + (z - z_1)^2}}{C_1 \cosh U_0} + C_2, \label{2.59}
\end{equation}

\begin{equation*}
    \omega_E = C_1 \frac{\sqrt{\rho^2 + (z - z_1)^2}}{(z - z_1) + \sqrt{\rho^2 + (z - z_1)^2} \cdot \tanh U_0}  \cdot \frac{1}{\cosh U_0} + C_3,
\end{equation*}

where $U_{0}, C_1$, $C_2$, and $C_3$ are arbitrary constants.

Straightforward calculations show  the solution (\ref{2.59}) turns all components of the Riemann-Christoffel curvature tensor to zero. Thus, it makes sense to call the solution (\ref{2.59}) a euclidon solution. One can say that it characterizes some relativistic noninertial frame of reference in flat space-time. Nevertheless, the solution (\ref{2.59}) allows one to generate solutions describing curved space-time.

In order to make a nonlinear "composition" of the solution (\ref{2.59}) with the seed solution, we use the method of variation of parameters. We shall consider $C_{1}$, $C_{2}$, $C_{3}$ and $U_{0}$ in the solution (\ref{2.59}) to be functions 

\begin{equation}
    C_1 \rightarrow f^0(\rho, z), \quad C_2 \rightarrow \omega^0(\rho, z), \quad C_3 \rightarrow \Phi^0(\rho, z), \quad U_0 \rightarrow U(\rho, z). \label{2.60}
\end{equation}

In this case we have

\begin{equation*}
    f = \frac{(z - z_1) + \sqrt{\rho^2 + (z - z_1)^2} \tanh U(\rho, z)}{f^0}, 
\end{equation*}

\begin{equation*}
    \Phi=\frac{\sqrt{\rho^2+(z-z_{1})^2}}{f^0\cdot \cosh U(\rho,z)}+\omega^0(\rho,z),
\end{equation*}

\begin{equation}
    \omega = f^0 \frac{\sqrt{\rho^2 + (z - z_1)^2}}{(z - z_1) + \sqrt{\rho^2 +
    (z - z_1)^2}\tanh U(\rho, z)}  \cdot \frac{1}{\cosh U(\rho, z)} + \label{2.61}
    \end{equation}
    
\begin{equation*}
    +\Phi^0(\rho, z).
\end{equation*}

A substitution of (\ref{2.61}) into Equations (\ref{2.58}) leads to the following set of first-order differential equations for the unknown function $U(\rho, z)$

\begin{equation*}
    \sqrt{\rho^2 + (z - z_1)^2} \frac{\partial U}{\partial \rho} = (z - z_1) \frac{1}{f^0} \frac{\partial f^0}{\partial \rho} + \frac{\rho}{f^0}\frac{\partial f^0}{\partial z}
    \end{equation*}
\begin{equation*}
    + \left[(z - z_1) \sinh U + \sqrt{\rho^2 + (z - z_1)^2} \cdot \cosh U \right] \frac{1}{f^0} \frac{\partial \Phi^0}{\partial \rho} + 
\end{equation*}

\begin{equation}
   + \rho \frac{\sinh U}{f^0}\frac{\partial \Phi^{0}}{\partial z}, \label{2.62}
\end{equation}

\begin{equation*}
    \sqrt{\rho^2 + (z - z_1)^2} \frac{\partial U}{\partial z} = -\frac{\rho}{f^0} \frac{\partial f^0}{\partial \rho}  +\frac{z-z_1}{f^0} \frac{\partial f^0}{\partial z}
    +
\end{equation*}

\begin{equation*}
   +  \left[ (z - z_1) \sinh U + \sqrt{\rho^2 + (z - z_1)^2} \cdot\cosh U \right] \frac{1}{f^0} \frac{\partial \Phi^0}{\partial z}-\rho\frac{\sinh U}{f^{0}}\frac{\partial \Phi^{0}}{\partial \rho}.
\end{equation*}

It is readily seen that the integrability condition for Equations (\ref{2.62}) is satisfied.

Equations  (\ref{2.62}) are nonlinear. The problem of linearization of the system (\ref{2.62}) can be solved by the substitution $U(\rho, z) = \ln [a(\rho, z)/b(\rho, z)]$. Then we have the following set of linear equations for the functions $a(\rho, z)$  and $b(\rho, z)$

\begin{equation*}
    2 f^0 \frac{\partial a}{\partial \rho} = a \left( \hat{L_1} + \frac{\partial}{\partial \rho} \right) f^0 - b \left( \hat{L_1} - \frac{\partial}{\partial \rho} \right) \Phi^0,
\end{equation*}

\begin{equation*}
   2 f^0 \frac{\partial a}{\partial z} = -a \left( \hat{L_2} - \frac{\partial}{\partial z} \right) f^0 + b \left( \hat{L_2} + \frac{\partial}{\partial z} \right) \Phi_0,
\end{equation*}
\begin{equation}
     2 f^0 \frac{\partial b}{\partial \rho} = -b \left( \hat{L_1} - \frac{\partial}{\partial \rho} \right) f^0 - a \left( \hat{L_1} + \frac{\partial}{\partial \rho} \right) \Phi^0,
\end{equation} \label{2.63}

\begin{equation*}
     2 f^0 \frac{\partial b}{\partial z} = b \left( \hat{L_2} + \frac{\partial}{\partial z} \right) f^0 + a \left( \hat{L_2} - \frac{\partial}{\partial z} \right) \Phi^0,
\end{equation*}

 where $\hat{L_1}$ and $\hat{L_2}$ are the linear operators

 \begin{equation*}
    \sqrt{\rho^2 + (z - z_1)^2} \hat{L_1} = (z - z_1) \frac{\partial}{\partial \rho} + \rho \frac{\partial}{\partial z},
\end{equation*}

 \begin{equation*}
    \sqrt{\rho^2 + (z - z_1)^2} \hat{L_2} = \rho \frac{\partial}{\partial \rho} - (z - z_1) \frac{\partial}{\partial z}.
\end{equation*}

Thus the problem of nonlinear superposition of the solution (\ref{2.61}) with an arbirary axially symmetric stationary solution of the Einstein equations reduces to integrating a set of first-order linear differential equations. Methods of treating this problem are well studied.

\section{Application of the euclidon method. Two-stationary euclidon solution (the Kerr-NUT solution)}

\textbf{1.} It is convenient to go over to prolate ellipsoidal coordinates  $(x, y)$ , which are connected to the Weyl coordinates $(\rho, z)$ by the relations 

\begin{equation}
    \rho=k_{0}\sqrt{(x^2-1)(1-y^2)}, \quad z=k_{0}xy \label{5.1}
\end{equation}

($k_{0}$ - a real constant). 

In terms of the coordinates $(x, y)$ the Equations (\ref{2.58}), (\ref{2.61}) and (\ref{2.61}) are rewritten

\begin{equation}
    \frac{\partial \omega^0}{\partial x} = \frac{k_0(1 - y^2)}{(f^0)^2} \frac{\partial \Phi^0}{\partial y}, \quad
    \frac{\partial \omega^0}{\partial y} = -\frac{k_0(x^2 - 1)}{(f^0)^2} \frac{\partial \Phi^0}{\partial x}, \label{5.2}
\end{equation}

\begin{equation}
    f = \frac{k_{0}}{f^0} \left[ xy - 1 + (x - y) \tanh U \right], \quad
    \Phi = \frac{k_{0}}{f^0} \frac{(x - y)}{\cosh U} + \omega^0, \label{5.3}
\end{equation} 

\begin{equation*}
    \omega =  f^0 \frac{(x - y)}{xy - 1 + (x - y) \tanh U} \frac{1}{\cosh U} + \Phi^0
\end{equation*}

 and

\begin{equation*}
    \frac{\partial U}{\partial x} = (\frac{xy - 1}{x - y}) \frac{1}{f^0} \frac{\partial f^0}{\partial x} + (\frac{1 - y^2}{x - y}) \frac{1}{f^0} \frac{\partial f^0}{\partial y}
\end{equation*}

\begin{equation*}
    + [\cosh U + (\frac{xy - 1}{x - y}) \sinh U] \frac{1}{f^0} \frac{\partial \Phi^0}{\partial x} + (\frac{1 - y^2}{x - y}) \frac{\sinh U}{f^{0}} \frac{\partial \Phi^0}{\partial y} ,
\end{equation*}

\begin{equation}
    \frac{\partial U}{\partial y} = -(\frac{x^2 - 1}{x - y}) \frac{1}{f^0} \frac{\partial f^0}{\partial x} + (\frac{xy-1}{x - y}) \frac{1}{f^0} \frac{\partial f^0}{\partial y} \label{5.4}
\end{equation}

\begin{equation*}
    + [\cosh U + (\frac{xy - 1}{x - y}) \sinh U] \frac{1}{f^0} \frac{\partial \Phi^0}{\partial y} - (\frac{x^2-1}{x - y}) \frac{\sinh U}{f^{0}} \frac{\partial \Phi^0}{\partial x}.
\end{equation*}

Choosing the function in the form of

\begin{equation}
    U=\ln\frac{a}{\sqrt{f^{0}}}-\ln \frac{b}{\sqrt{f^{0}}} \label{5.5}
\end{equation}

We arrive at the following system of linear differential equations for the functions

\begin{equation*}
  2f^{0}\frac{\partial a}{\partial x}=a(1+y)\hat{L_1}f^{0}+b(1-y)\hat{L_2}\Phi^{0},  
\end{equation*}

\begin{equation*}
  2f^{0}\frac{\partial a}{\partial y}=-a(x-1)\hat{L_2}f^{0}+b(x+1)\hat{L_1}\Phi^{0},  
\end{equation*}

\begin{equation}
  2f^{0}\frac{\partial b}{\partial x} = b(1-y)\hat{L}_2f^{0} - a(1+y)\hat{L}_1\Phi^{0}, \label{5.6}
\end{equation}

\begin{equation*}
  2f^{0}\frac{\partial b}{\partial y}=b(x+1)\hat{L_1}f^{0}+a(x-1)\hat{L_2}\Phi^{0}.  
\end{equation*}

The operators take the form

\begin{equation*}
    (x-y)\hat{L_1}=(x-1)\frac{\partial}{\partial x}+(1-y)\frac{\partial}{\partial y},
\end{equation*}

\begin{equation*}
    (x-y)\hat{L_2}=(x+1)\frac{\partial}{\partial x}-(1+y)\frac{\partial}{\partial y}.
\end{equation*}

 \textbf{1-I.}
 If we choose 
\begin{equation}
   f^0=1, \quad \Phi^0 = \omega^0=0, \label{5.7}
\end{equation}

 in this case we obtain from \ref{5.3}, \ref{5.4}

 \begin{equation}
   f=f^{0}_{1,-} = k_{0}(xy-1) + k_{0}(x - y) \tanh U_0, \quad \Phi = \Phi^{0}_{1,-}=\frac{k_{0}(x - y)}{\cosh U_0}, \label{5.8}
\end{equation}

\begin{equation*}
    \omega =   \frac{(x - y)}{xy - 1 + (x - y) \tanh U_{0}} \frac{1}{\cosh U_{0}} ,
\end{equation*}

where  $U=U_{1,-}=U_{0}$.
 
 \textbf{1-II.}
 If we choose
\begin{equation}
   f^0=f^{0}_{1,-} , \quad \Phi^0 = \Phi^{0}_{1,-}, \label{5.9}
\end{equation}
in this case we obtain from \ref{5.3}, \ref{5.4}

\begin{equation*}
    f = f^0_{2,-,-}= \frac{k_{o}(xy - 1) + k_{o}(x - y) \tanh U}{f^{0}_{1,-}},\quad \Phi =\Phi^0_{2,-,-} = \frac{k_{0}}{f^0_{1,-}} \frac{(x - y)}{\cosh U} + \omega^0_{1,-} ,
\end{equation*}

where $U=U_{2,-,-}=\ln\frac{a}{b}= const$,
\begin{equation*}
   \frac{a}{b}=-\sqrt{\frac{1+\tanh U_0}{1-\tanh U_0}}, \quad \tanh U=\tanh U_0 , \quad \cosh U=-\cosh U_0,
\end{equation*}

 and 

 \begin{equation}
   f=f^0_{2,-,-}=1, \quad \Phi = \omega=0. \label{5.10}
\end{equation}

 \textbf{1-III.} Two-stationary euclidon solution.

If we choose the external field in the form

 \begin{equation}
    f^0 = f^0_{1,+}=k_{0}(xy + 1) + k_{0}(x + y) \tanh U_0, \quad \Phi^0=\Phi^0_{1,+} = \frac{k_{0}(x + y)}{\cosh U_0}, \label{5.11}
\end{equation}

\begin{equation*}
   \omega^0=\omega^0_{1,+}=\frac{(x+y)}{xy+1+(x+y)\cdot\tanh U_{0}}\cdot\frac{1}{\cosh U_{0}}
\end{equation*}

 where $U_0$ is a constant , we find

\begin{equation*}
    \hat{L}_1 f^0 =k_{0}( 1 + \tanh U_0), \quad \hat{L}_2 \Phi^0 = \frac{k_{0}}{\cosh U_0},
\end{equation*}

\begin{equation*}
    \hat{L}_2 f^0 = k_{0}(\tanh U_0 - 1), \quad \hat{L}_1 \Phi^0 = \frac{ k_{0}}{\cosh U_0}
\end{equation*}

\begin{equation*}
    k^{-1}_{0} a = (x + 1) \sqrt{1 + \tanh U_0} - \alpha_0 (1 - y) \sqrt{1 - \tanh U_0},
\end{equation*}

\begin{equation*}
    k^{-1}_{0}b = \alpha_0 (1 + y) \sqrt{1 + \tanh U_0} - (x - 1) \sqrt{1 - \tanh U_0}.
\end{equation*}

Here $\alpha_0$ is an integration constant. The function $U$ in this case has the form \cite{162}

\begin{equation*}
   U(x,y)= U_{2,-,+}(x,y) = \ln[  (x + 1) \sqrt{1 + \tanh U_0} - c_0 (1 - y) \sqrt{1 - \tanh U_0}]-
\end{equation*}
\begin{equation}
    -\ln[ c_{0} (1+y) \sqrt{1 + \tanh U_0} -  (x-1) \sqrt{1 - \tanh U_0}]. \label{5.12}
\end{equation}

If we set in  

\begin{equation}
    \tanh U_0 =\tanh U^a_0= \frac{1 - a_0^2}{1 + a_0^2}, \quad c_0 = \frac{b_0 - a_0}{1 + a_0 b_0}, \label{5.13}
\end{equation}

then, accordingly, a stationary two-Euclidean solution is constructed, which coincides with the Kerr-NUT solution in form:

\begin{equation*}
    f = f^0_{2,-,+}=\frac{k_{o}(xy - 1) + k_{o}(x - y) \tanh U(x,y,a_{0},b_{0})}{f^0_{1,+}}=\frac{A_2}{B_2}, 
    \end{equation*}

    \begin{equation*}
    \quad \Phi =\Phi^0_{2,-,+} = \frac{k_{0}}{f^0_{1,+}} \cdot \frac{(x - y)}{\cosh U(x,y,a_{0},b_{0})} + \omega^0_{1,+}= \frac{C_2}{B_2},
\end{equation*}

\begin{equation*}
    A_2 = (x^2 - 1)(1 + a_0 b_0)^2 + (y^2 - 1)(b_0 - a_0)^2,
\end{equation*}

\begin{equation}
    B_2 = \left[ (1 + a_0 b_0) x + (1 - a_0 b_0) \right]^2 + \left[ (b_0 - a_0) y + (b_0 + a_0) \right]^2, \label{5.14}
\end{equation}

\begin{equation*}
    C_2 = 2 \left[ (b_0 + a_0) (1 + a_0 b_0) x - (b_0 - a_0) (1 - a_0 b_0) y \right].
\end{equation*}

 If we put 

\begin{equation*}
    \frac{1+a_{0}b_{0}}{\sqrt{(1+a^2_{0})(1+b^2_{0})}}=k_{0}(m^2_{0}+l^2_{0})^{-1/2}, \frac{b_{0}-a_{0}}{\sqrt{(1+a^2_{0})(1+b^2_{0})}}=a(m^2_{0}+l^2_{0})^{-1/2},
\end{equation*}

\begin{equation*}
    \frac{1-a_{0}b_{0}}{\sqrt{(1+a^2_{0})(1+b^2_{0})}}=m_{0}(m^2_{0}+l^2_{0})^{-1/2}, \frac{b_{0}+a_{0}}{\sqrt{(1+a^2_{0})(1+b^2_{0})}}=-l_{0}(m^2_{0}+l^2_{0})^{-1/2}, \label{2.93}
\end{equation*}
 we can obtain the Kerr-NUT   solution in Boyer-Lindquist coordinates $(r,\theta) \ \  (x=(r-m_{0})/k_{0}, \ \ y= cos\theta)$ 

\begin{equation*}
    ds^2=\frac{r^2+(a\cos \theta-l_{0})^2}{r^2-2m_{0}r+a^2-l^2_{0}}dr^2+[r^2+(a\cos\theta-l_{0})^2]\times
\end{equation*}

\begin{equation*}
    \times[d\theta^2+\frac{(r^2-2m_{0}r+a^2-l^2_{0})\sin^2 \theta d\varphi^2}{r^2-2m_{0}r+a^2\cos^2 \theta -l^2_{0}}]-[1-2\frac{m_{0}r+l_{0}(l_{0}-a\cos\theta)}{r^2+(a\cos \theta-l_{0})^2}]\times
\end{equation*}

\begin{equation*}
    \times\{dt-(\frac{2a\sin^2 \theta[m_{0}r+l_{0}(l_{0}-a\cos \theta)]}{r^2-2m_{0}r+a^2\cos^2\theta-l^2_{0}}-2l_{0}\cos \theta)d\varphi\}^2. 
\end{equation*}

If we set $a_0 = -b_0$, then (\ref{5.14}) yields the Kerr solution.

\textbf{2-IV.} 


\begin{equation*}
    f = f^0_{2,+,-}= \frac{k_{o}(xy + 1) + k_{o}(x + y) \tanh U}{f^0_{1,-}}, 
\end{equation*}

\begin{equation}
    \quad \Phi =\Phi^0_{2,+,-} = \frac{k_{0}}{f^0_{1,-}} \frac{(x + y)}{\cosh U} + \omega^0_{1,-},
\end{equation} \label{5.15}

where
    \begin{equation*}
   U=U_{2,+,-}(x,y) = U_{2,-,+}(-x,y).
   \end{equation*}

\section{Superposition of two-stationary euclidon solution with an arbitrary stationary vacuum field}

 Nonlinear superposition of the Kerr solution with an arbitrary axially symmetric Einstein field can be fulfilled by the variation of constants in the solution obtainable from the Kerr one by the constant phase transformation \cite{158}, \cite{159}, \cite{160}. We shall carry out the consideration in terms of the functions $f$ and $\Phi$ satisfying, as is known, the equations 

 \begin{equation}
    f \Delta f = (\overrightarrow{\nabla} f)^2 - (\overrightarrow{\nabla} \Phi)^2, \quad f \Delta \Phi = 2 \overrightarrow{\nabla} f  \overrightarrow{\nabla} \Phi. \label{6.1}
\end{equation}

To perform the superposition of the two-stationary Kerr solution with an arbitrary solution of the system (\ref{6.1}) it is not enough to variate one constant because both the functions $f^{0} $  and $\Phi^{0} $  are not equal to zero. Therefore, we should take as "seed" variating solution not the  Kerr solution  but its generalization containing two independent constants $a_{0}$ and $b_{0}$, e.g. the solution of the form (\ref{5.14})

\begin{equation*}
    \frac{f - 1}{1} = -\frac{2[x(1 - a_0^2 b_0^2) + y(b_0^2 - a_0^2) + (a_0 + b_0)^2 + (a_0 b_0 - 1)^2]}{[x(1 + a_0 b_0) + 1 - a_0 b_0]^2 + [y(b_0 - a_0) + a_0 + b_0]^2},
\end{equation*}

\begin{equation}
    \frac{\Phi - 0}{1} = \frac{2[x(a_0 + b_0)(1 + a_0 b_0) + y(a_0 - b_0)(1 - a_0 b_0)]}{[x(1 + a_0 b_0) + 1 - a_0 b_0]^2 + [y(b_0 - a_0) + a_0 + b_0]^2}. \label{6.2}
\end{equation} 

which reduces to Kerr solution  bu at  at $b_{0}=-a_{0}.$

Then the superposition formulae generalizing (\ref{6.2}) in the case $\Phi^{0}\neq0$ take the form 

\begin{equation*}
    \frac{f - f^0}{f^0} = -\frac{2[x(1 - a^2 b^2) + y(b^2 - a^2) + (a + b)^2 + (1 - ab)^2]}{[x(1 + ab) + 1 - ab]^2 + [y(b - a) + a + b]^2},
\end{equation*}

\begin{equation}
    \frac{\Phi - \Phi^0}{f^0} = \frac{2[x(a + b)(1 + ab) + y(a - b)(1 - ab)]}{[x(1 + ab) + 1 - ab]^2 + [y(b - a) + a + b]^2}, \label{6.3}
\end{equation}

where $a$ and $b$ are some functions of the coordinates $(x, y)$ and should be determined from the field equations. 

Substituting (\ref{6.3}) into (\ref{6.1}) we can find the dependence of $a$ and $b$ on the functions $f^{0} $  and $ \Phi^{0} $: 

\begin{equation*}
    a,_x = -\frac{1}{2(x + y) f^0} \{ [(xy + 1) f^0,_x + (1 - y^2) f^0,_y] 2a 
\end{equation*}
\begin{equation*}
    + [(xy + 1) \Phi^0,_x + (1 - y^2) \Phi^0,_y] (1 - a^2) + (x + y)(1 + a^2) \Phi^0,_x \},
\end{equation*}

\begin{equation*}
    a,_y = -\frac{1}{2(x + y) f^0} \{ [-(x^2 - 1) f^0,_x + (xy + 1) f^0,_y] 2a 
\end{equation*}

\begin{equation}
    + [-(x^2 - 1) \Phi^0,_x + (xy + 1) \Phi^0,_y] (1 - a^2) + (x + y)(1 + a^2) \Phi^0,_y \}, \label{6.4}
\end{equation}

\begin{equation*}
    b,_x = \frac{1}{2(x - y) f^0} \{ [(xy - 1) f^0,_x + (1 - y^2) f^0,_y] 2b 
\end{equation*}

\begin{equation*}
    + [(xy - 1) \Phi^0,_x + (1 - y^2) \Phi^0,_y] (1 - b^2) - (x - y)(1 + b^2) \Phi^0,_x \},
\end{equation*}

\begin{equation*}
    b,_y = \frac{1}{2(x - y) f^0} \{ [-(x^2 - 1) f^0,_x + (xy - 1) f^0,_y] 2b 
\end{equation*}

\begin{equation*}
    + [-(x^2 - 1) \Phi^0,_x + (xy - 1) \Phi^0,_y] (1 - b^2) - (x - y)(1 + b^2) \Phi^0,_y \}.
\end{equation*}

Relations (\ref{6.3}) and (\ref{6.4}) determine the desired superposition of the solution (\ref{6.2}) with an arbitrary Einstein vacuum field. 

If we put

\begin{equation}
    a = \exp(-U^a), \quad b = -\exp U^b. \label{6.5}
\end{equation}

from (\ref{6.4}) we obtain 

\begin{equation*}
    \frac{\partial U^a}{\partial x} = (\frac{xy + 1}{x + y}) \frac{1}{f_0} \frac{\partial f_0}{\partial x} + (\frac{1 - y^2}{x + y}) \frac{1}{f_0} \frac{\partial f_0}{\partial y}
\end{equation*}

\begin{equation*}
    + [\cosh U^a + (\frac{xy + 1}{x + y}) \sinh U^a] \frac{1}{f_0} \frac{\partial \Phi_0}{\partial x} + (\frac{1 - y^2}{x + y}) \frac{\sinh U^a}{f_{0}} \frac{\partial \Phi_0}{\partial y} ,
\end{equation*}

\begin{equation}
    \frac{\partial U^a}{\partial y} = -(\frac{x^2 - 1}{x + y}) \frac{1}{f_0} \frac{\partial f_0}{\partial x} + (\frac{xy+1}{x + y}) \frac{1}{f_0} \frac{\partial f_0}{\partial y} \label{6.6}
\end{equation}

\begin{equation*}
    + [\cosh U^a + (\frac{xy + 1}{x + y}) \sinh U^a] \frac{1}{f_0} \frac{\partial \Phi_0}{\partial y} - (\frac{x^2-1}{x + y}) \frac{\sinh U^a}{f_{0}} \frac{\partial \Phi_0}{\partial x}.
\end{equation*}

and

\begin{equation*}
    \frac{\partial U^b}{\partial x} = (\frac{xy - 1}{x - y}) \frac{1}{f_0} \frac{\partial f_0}{\partial x} + (\frac{1 - y^2}{x - y}) \frac{1}{f_0} \frac{\partial f_0}{\partial y}
\end{equation*}

\begin{equation*}
    + [\cosh U^b + (\frac{xy - 1}{x - y}) \sinh U^b] \frac{1}{f_0} \frac{\partial \Phi_0}{\partial x} + (\frac{1 - y^2}{x - y}) \frac{\sinh U^b}{f_{0}} \frac{\partial \Phi_0}{\partial y} ,
\end{equation*}

\begin{equation}
    \frac{\partial U^b}{\partial y} = -(\frac{x^2 - 1}{x - y}) \frac{1}{f_0} \frac{\partial f_0}{\partial x} + (\frac{xy-1}{x - y}) \frac{1}{f_0} \frac{\partial f_0}{\partial y} \label{6.7}
\end{equation}

\begin{equation*}
    + [\cosh U^b + (\frac{xy - 1}{x - y}) \sinh U^b] \frac{1}{f_0} \frac{\partial \Phi_0}{\partial y} - (\frac{x^2-1}{x - y}) \frac{\sinh U^b}{f_{0}} \frac{\partial \Phi_0}{\partial x}.
\end{equation*}

In this way, many generalizations of the Kerr solution describing axially symmetric rotating masses have been obtained \cite{1}. 
\section{Three- und four- stationary euclidon solution}
\textbf{1-I.} Let us consider superposition (\ref{6.3}) of the stationary two-euclidon solution (\ref{6.2}) whith the stationary own-euclidon solution $(f^{0}_{1,-},\ \ \Phi^{0}_{1,-})$.

This  stationary three-euclidon solution is easy to rewrite as

\begin{equation}
    f=f^0_{3,-,+,-} =\frac{k_{0}(xy - 1) + k_{0}(x - y) \tanh U(a,b)}{f^0_{2,+,-}}, \label{8.1}
    \end{equation}

    \begin{equation}
    \Phi=\Phi^0_{3,-,+,-} =\frac{1}{f^0_{2,+,-}}\cdot\frac{k_{0}(x - y)}{\cosh U(a,b)} +\omega^0_{2,+,-}, \label{8.2}
\end{equation}

where
\begin{equation}
    f^0_{2,+,-} =\frac{k_{0}(xy + 1) + k_{0}(x + y) \tanh U^a}{f^0_{1,-}}, 
    \end{equation}

\begin{equation*}
   \omega^0_{2,+,-}=f^0_{1,-}\frac{(x+y)}{xy+1+(x+y)\cdot\tanh U^a}\cdot\frac{1}{\cosh U^a}+ \Phi^{0}_{1,-}.
\end{equation*}

And from (\ref{5.12}), (\ref{6.5}), (\ref{6.6}) and (\ref{6.7})

\begin{equation*}
   U(a,b)= U_{3,-,+,-}(x,y)= 
   \end{equation*}
   \begin{equation*}
   = \ln[  (x + 1) \sqrt{1 + \tanh U^a} - \alpha (1 - y) \sqrt{1 - \tanh U^a}]-
\end{equation*}
\begin{equation}
    -\ln[ \alpha (1+y) \sqrt{1 + \tanh U^a} -  (x-1) \sqrt{1 - \tanh U^a}], \label{7.4}
\end{equation}

where
\begin{equation}
     \tanh U^a=\frac{1 - a^2}{1 + a^2}, \quad \alpha = \frac{b - a}{1 + a b}, \label{7.5}
\end{equation}

\begin{equation*}
    a = \exp(-U^a), \quad b = -\exp(U^b).
\end{equation*}

\begin{equation*}
    U^a=U_{2,+,-}(x,y), \ \  U^b=U_{2,-,-}(x,y).
 \end{equation*}

From (\ref{5.3}) it is easy to see that these formulas also define a superposition  of the stationary own-euclidon solution (\ref{5.8}) whith the stationary two-euclidon solution $(f^0_{2,+,-}, \quad \omega^0_{2,+,-})$ with $U=U(a,b)= U_{3,-,+,-}(x,y)$.

\textbf{1-II.}

\begin{equation*}
    f=f^0_{3,-,-,+} =\frac{k_{0}(xy - 1) + k_{0}(x - y) \tanh U(a,b)}{f^0_{2,-,+}}, 
    \end{equation*}

    \begin{equation}
    \Phi=\Phi^0_{3,-,-,+} =\frac{1}{f^0_{2,-,+}}\cdot\frac{k_{0}(x - y)}{\cosh U(a,b)} +\omega^0_{2,-,+},\label{7.6}
\end{equation}

where
\begin{equation*}
    f^0_{2,-,+} =\frac{k_{0}(xy - 1) + k_{0}(x - y) \tanh U^a}{f^0_{1,+}}, 
    \end{equation*}

\begin{equation}
   \omega^0_{2,-,+}=f^0_{1,+}\frac{(x-y)}{xy-1+(x-y)\cdot\tanh U^a}\cdot\frac{1}{\cosh U^a}+ \Phi^{0}_{1,+}. \label{7.7}
\end{equation}

And from (\ref{5.12}), (\ref{6.5}), (\ref{6.6}) and (\ref{6.7})

\begin{equation*}
   U(a,b)= U_{3,-,-,+}(x,y)= 
   \end{equation*}
   \begin{equation*}
   = \ln[  (x + 1) \sqrt{1 + \tanh U^a} - \alpha (1 - y) \sqrt{1 - \tanh U^a}]-
\end{equation*}
\begin{equation}
    -\ln[ \alpha (1+y) \sqrt{1 + \tanh U^a} -  (x-1) \sqrt{1 - \tanh U^a}], \label{7.8}
\end{equation}

where
\begin{equation}
     \tanh U^a=\tanh U_{0} =\frac{1 - a^2}{1 + a^2}, \quad \alpha = \frac{ a-b }{1 + a b}, \label{7.9}
\end{equation}

\begin{equation*}
    U^a=U_{2,+,+}(x,y)=U_{2,-,-}(x,y), \ \  U^b= U_{2,-,+}(x,y),
 \end{equation*}

 \begin{equation*}
    a=\exp{(-U^a)}=-\sqrt{\frac{1-\tanh U_{0}}{1+\tanh U_{0}}},
 \end{equation*}

 \begin{equation*}
      b=-\exp{U^b}=-\frac{(x + 1) \sqrt{1 + \tanh U_{0}} - \alpha_{0} (1 - y) \sqrt{1 - \tanh U_{0}}}{\alpha_{0} (1+y) \sqrt{1 + \tanh U_{0}} -  (x-1) \sqrt{1 - \tanh U_{0}}},
 \end{equation*}

 If $\alpha_{0}\rightarrow \infty$ we have
\begin{equation*}
    \alpha=-\frac{\sqrt{1 - \tanh^2 U_{0}}}{y+\tanh U_{0}}
 \end{equation*}

 \begin{equation*}
    \exp {U(a,b)}=-\sqrt{\frac{1+\tanh U_{0}}{1-\tanh U_{0}}}, \ \ \tanh U(a,b) =\tanh U_{0}
 \end{equation*}

and
\begin{equation*}
    f^0_{3,-,-,+} \rightarrow {f^0_{1,+}}, 
    \end{equation*}
    
 \begin{equation}
    \Phi^0_{3,-,-,+} \rightarrow \Phi^0_{1,+}. \label{7.10}
\end{equation}

Similarly for $\alpha_{0}\rightarrow 0$.

\textbf{2.}

Superposition of the stationary two-euclidon solution (\ref{6.2}) whith the stationary two-euclidon solution $(f^{0}_{2,-,+},\ \ \Phi^{0}_{2,-,+})$ gives stationary four-euclidon solution

\begin{equation*}
    f=f^0_{4,-,+,-,+} =\frac{k_{0}(xy - 1) + k_{0}(x - y) \tanh U(a,b)}{f^0_{3,+,-,+}}, 
    \end{equation*}

    \begin{equation}
    \Phi=\Phi^0_{4,-,+,-,+} =\frac{1}{f^0_{3,+,-,+}}\cdot\frac{k_{0}(x - y)}{\cosh U(a,b)} +\omega^0_{3,+,-,+}, \label{7.11}
\end{equation}

where
\begin{equation*}
    f^0_{3,+,-,+}(x,y)=-f^0_{3,-,+,-}(-x,y), 
    \end{equation*}

\begin{equation}
   \omega^0_{3,+,-,+}(x,y)=-\omega^0_{3,-,+,-}(-x,y). \label{7.12}
\end{equation}

And 

\begin{equation*}
   U(a,b)= U_{4,-,+,-,+}(x,y)= 
   \end{equation*}
   \begin{equation*}
   = \ln[  (x + 1) \sqrt{1 + \tanh U^a} - \alpha (1 - y) \sqrt{1 - \tanh U^a}]-
\end{equation*}
\begin{equation}
    -\ln[ \alpha (1+y) \sqrt{1 + \tanh U^a} -  (x-1) \sqrt{1 - \tanh U^a}], \label{7.13}
\end{equation}

where
\begin{equation*}
     \tanh U^a=\frac{1 - a^2}{1 + a^2}, \quad \alpha = \frac{b - a}{1 + a b},
\end{equation*}

\begin{equation*}
    a=\exp{(-U^a)}, \ \  b=-\exp{U^b},
\end{equation*}

\begin{equation*}
    U^a=U_{3,+,-,+}(x,y), \ \  U^b=U_{3,-,-,+}(x,y).
 \end{equation*}

From (\ref{7.11}) it is easy to see that these formulas also define a superposition  of the stationary own-euclidon solution (\ref{5.8}) whith the stationary three-euclidon solution $(f^0_{3,+,-,+}, \quad \omega^0_{3,+,-,+})$ with $U=U(a,b)= U_{4,-,+,-,+}(x,y)$.

 In a similar way we can get  the stationary 2N-euclidon solution containing N-Kerr solution in which all gravitating centers coincide with each other and reducing to the Zipoy solution \cite{163} in a static limit. 

\section{ Euclidon algebra}

From \ref{2.61}, \ref{2.62} follows the law of composition of two euclidons
\begin{equation}
    f_{E,i} \times f^{-1}_{E,k}= 
\begin{pmatrix}
f^{-1}_{1,k}(U_{1,k})f_{1,i}(U_{2,i,k})\\
f^{-1}_{1,k}(U_{1,k})\Phi_{1,i}(U_{2,i,k})+\omega_{1,k}(U_{1,k}) \\
f_{1,k}(U_{1,k})\omega_{1,k}(U_{2,i,k})+\Phi_{1,k}(U_{1,k}) 
\end{pmatrix} \label{9.1}
 \end{equation}

where
\begin{equation}
    f_{Ei} = 
\begin{pmatrix}
f_{1,i}(U_{1,i})\\
\Phi_{1,i}(U_{1,i}) \\
\omega_{1,i}(U_{1,i}) 
\end{pmatrix}; \\ i,k,l=\pm. \label{9.2}
 \end{equation}

From \ref{9.1}and \ref{5.10} follows
\begin{equation*}
    f_{E,i} \times f^{-1}_{E,i}= e
\end{equation*}
where single element
\begin{equation}
    e = 
\begin{pmatrix}
1\\
0 \\
0 
\end{pmatrix}. \label{9.3}
 \end{equation}

Based on \ref{9.1}, \ref{9.3} for the inverse element
\begin{equation}
    e\times f^{-1}_{E,i}= f^{-1}_{E,i}=
\begin{pmatrix}
f^{-1}_{1,i}(U_{1,i})\\
\omega_{1,i}(U_{1,i}) \\
\Phi_{1,i}(U_{1,i}) 
\end{pmatrix}. \\ \label{9.4}
 \end{equation}

 It's easy to see that
\begin{equation}
     (f^{-1}_{E,i})^{-1}= f_{E,i} \label{9.5}
\end{equation}
 and
 \begin{equation}
    (f_{E,i} \times f^{-1}_{E,k})^{-1}= f^{-1}_{E,i}\times f_{E,k}= f_{E,k}\times f^{-1}_{E,i}. \label{9.6}
\end{equation}
In \ref{9.6} it is taken into account that $f_{E,k}$ is an external field and the change of indices is not significant.

Using \ref{9.1}, \ref{9.6} expressions \ref{8.1}, \ref{8.2} can be obtained in the following two ways
\begin{equation*}
    (f_{E,i} \times f^{-1}_{E,k})\times f_{E,l}=  f_{E,i} \times (f_{E,k} \times f^{-1}_{E,l})^{-1}=
    \end{equation*}
\begin{equation}
    =
\begin{pmatrix}
f^{-1}_{1,k}(U_{2,k,l})f_{1,l}(U_{1,l})f_{1,i}(U_{3,i,k,l})\\
f^{-1}_{1,k}(U_{2,k,l})f_{1,l}(U_{1,l})\Phi_{1,i}(U_{3,i,k,l})+f_{1,l}(U_{1,l})\omega_{1,k}(U_{2,k,l})+\Phi_{1,l}(U_{1,l}) \\
f_{1,k}(U_{2,k,l})f^{-1}_{1,l}(U_{1,l})\omega_{1,i}(U_{3,i,k,l})+f^{-1}_{1,l}(U_{1,l})\Phi_{1,k}(U_{2,k,l})+\omega_{1,l}(U_{1,l}) 
\end{pmatrix}. \label{9.7}
\end{equation}

In \ref{9.7} it is taken into account that  $U_{3,i,(k,l)}=U_{3,(i,k),l}$.
\ref{9.7} given \ref{9.6} is the associativity condition.

Thus the set of k-euclidons  forms a group.
\newpage

\section{  Conclusion}

Thus, using the simple nonlinear superposition method  we got the stationary 2N-euclidon (soliton) solution containing N-Kerr solution in which all gravitating centers coincide with each other and reducing to the Zipoy solution \cite{163} in a static limit.

Using Bonnor's theorem \cite{164} and Euclidon algebra, it is easy to construct a solution describing N massive magnetic dipoles in which all gravitating centers coincide with each other and reducing to the Zipoy solution \cite{1} in a static limit.

Based on 2N-euclidon solution in \cite{165} proposed a novel class of KERR-SEN solutions respecting the SO(2) symmetry group, systematically constructed via the Laurent series expansion technique. 

Developments in string theory and low-energy effective actions for supergravity provided new motivation to consider additional fields—such as the dilaton and the
Kalb-Ramond (axion) field—beyond the traditional gravitational field. Sen’s solution \cite{166} highlighted the importance of including these scalar fields for a more unified description of black holes in a string-theoretic
framework. These models predict that rotating solutions acquire modifications in their
horizon structure, global charges, and thermodynamic properties when coupled to scalar
(and pseudoscalar) fields.

Based on these considerations in \cite{165} received the multiplet $(f,\Phi, \phi , \chi) $, 
where $ \phi  $ and $ \chi $ are dilaton and axion.

Against this background, the KERR-SEN class of solutions emerges as a valuable
intersection of classical and quantum-corrected perspectives in gravity. On the one hand, the KERR-SEN solutions preserve the key stationary and axisymmetric features of the Kerr (and Kerr-NUT) metrics, ensuring their relevance for describing astrophysical black holes. On the other hand, by incorporating additional fields, they admit a wider range of mass, angular momentum, electric/magnetic charge, and scalar charges—potentially shedding light on how compact objects might deviate from the predictions of pure General
Relativity in extreme regimes.

\end{document}